\documentstyle[11pt,newpasp,twoside,epsf]{article}
\def\bge{\begin{equation}}
\def\ede{\end{equation}}
\def\ut#1{\mathop{\vtop{\ialign{##\crcr
     $\hfil\displaystyle{#1}\hfil$\crcr\noalign
     {\kern1pt\nointerlineskip}\hbox{$\hfil\sim\hfil$}\crcr
     \noalign{\kern1pt}}}}}

\def\edcomment#1{\iffalse\marginpar{\raggedright\sl#1\/}\else\relax\fi}
\reversemarginpar

\begin{document}
\title{The contribution of primordial globular clusters to the central galactic 
activity}
\author{R. Capuzzo--Dolcetta}
\affil{University of Rome ``La Sapienza'', Physics Dept., P.le A. Moro, 2 -- 
I-00185 -- Rome (Italy)}
\begin{abstract}
Globular clusters in galaxies have a mutual feedback with the environment, which 
is tuned by their dynamical evolution. This feedback may be the explanation of 
various features of both the globular cluster system  and the host galaxy. 
Relevant examples of these are the radial distribution of globulars and the 
violent initial activity of galaxies as AGN. 
\end{abstract}

\section{Introduction}
The existence of a strong connection between globular clusters (GCs) and their 
host galaxy is, nowadays, clear. Let us cite, for eample, correlations, like 
that between the mean metallicity of the globular cluster system (GCS) and 
galaxy luminosity (van den Bergh 1975; Brodie \& Huchra 1991; Durrel et al. 
1996). It resembles the positive correlation between galaxy stellar metallicity 
and galaxy luminosity ($Z \propto L^{0.4}$). In this context, Forbes \& Forte 
(2001) find, actually, that a correlation between GC colour and galaxy velocity 
dispersion exists just for the red subpopulation of GCs in a set of 28 early-
type galaxies showing a bimodal GC colour distribution (colour bimodality of GC 
population is a quite common characteristic of giant elliptical galaxies, and it 
has been found in low-luminosity ellipticals, also as shown by Forte et al. 
2001). Another, different, feature is that shown by GCs in NGC 1316 where red 
clusters show a strong correlation with the galaxy elongation while the blue 
ones are circularly distributed, being the two population equally concentrated 
(G\'omez et al. 2001).
These, and other, examples (see Harris 2000 for a review) are indicating that 
the link between GCSs and host galaxies is significant; by the way, this link 
may be both a result of the initial formation processes and/or due to further 
evolution and feedback.
\par Actually, evolution of the GCS in a galaxy is expected, due to that clusters 
suffer along their orbits of frictional braking and tidal torques, these latter 
exerted by both the galaxy large scale mass distribution and, on smaller scales, 
by the disk (in spiral galaxies) and the nucleus (often a super massive black 
hole). We refer to Capuzzo--Dolcetta (1993) for a discussion of these physical 
phenomena.
The dynamical evolution is not just changing the spatial distribution of GCs in 
the galaxy and the distribution of some characteristic parameters of the GCS, 
but has also {\it large-scale} (global) and {\it small-scale} (local) effects on 
the host galaxy, corresponding to a clear feedback between GCS and parent 
galaxy. For instance, globular cluster disruption contributes to the stellar 
bulge population on a large scale and to the nuclear population on a small 
scale. Observed tracers of dynamical interaction are the observed tidal tails of 
GCs that is a quite common feature (see Leon et al. 2000).
To find observational evidence of the nucleus-GC interaction is, of course, much 
more difficult. Probably, a better investigation of the photometric and 
kinematic characteristics of nuclear star clusters will be helpful in this
(B\"oker et al. 2000).
The topic of this report is to discuss one of these small-scale effects, namely 
that corresponding to the fate of globular clusters that have dissipated their 
orbital energy (and angular momentum) so to be confined in the nuclear galactic 
region where they lose stars to the external field that may be dominated by a 
compact massive black hole. Computations show that this loss of matter occurs on 
a time scale and in a quantity such to give an explanation of both the growing 
of the central galatic nucleus and of its activity as AGN.

\section{The formation of a supercluster at the galactic centre that feeds the 
central black hole}
Various papers (McLaughlin 1995; Capuzzo--Dolcetta \& Vignola 1997; Capuzzo
--Dolcetta \& Tesseri 1999, Capuzzo--Dolcetta \& Donnarumma 2001) have dealt with 
the topic of determining the quantity of stellar mass lost by the GCS of a 
galaxy. This can be done in the hypothesis that the presently observed 
difference between the GCS and the stellar field radial profiles is consequence 
of evolution from an initial state when the two components had the same 
concentration. Due to effects as dynamical friction and tidal interaction with 
the galaxy, the GCS initial distribution should, indeed, have evolved 
significantly (see left panel of Fig. 1). 

\begin{figure}
\plottwo{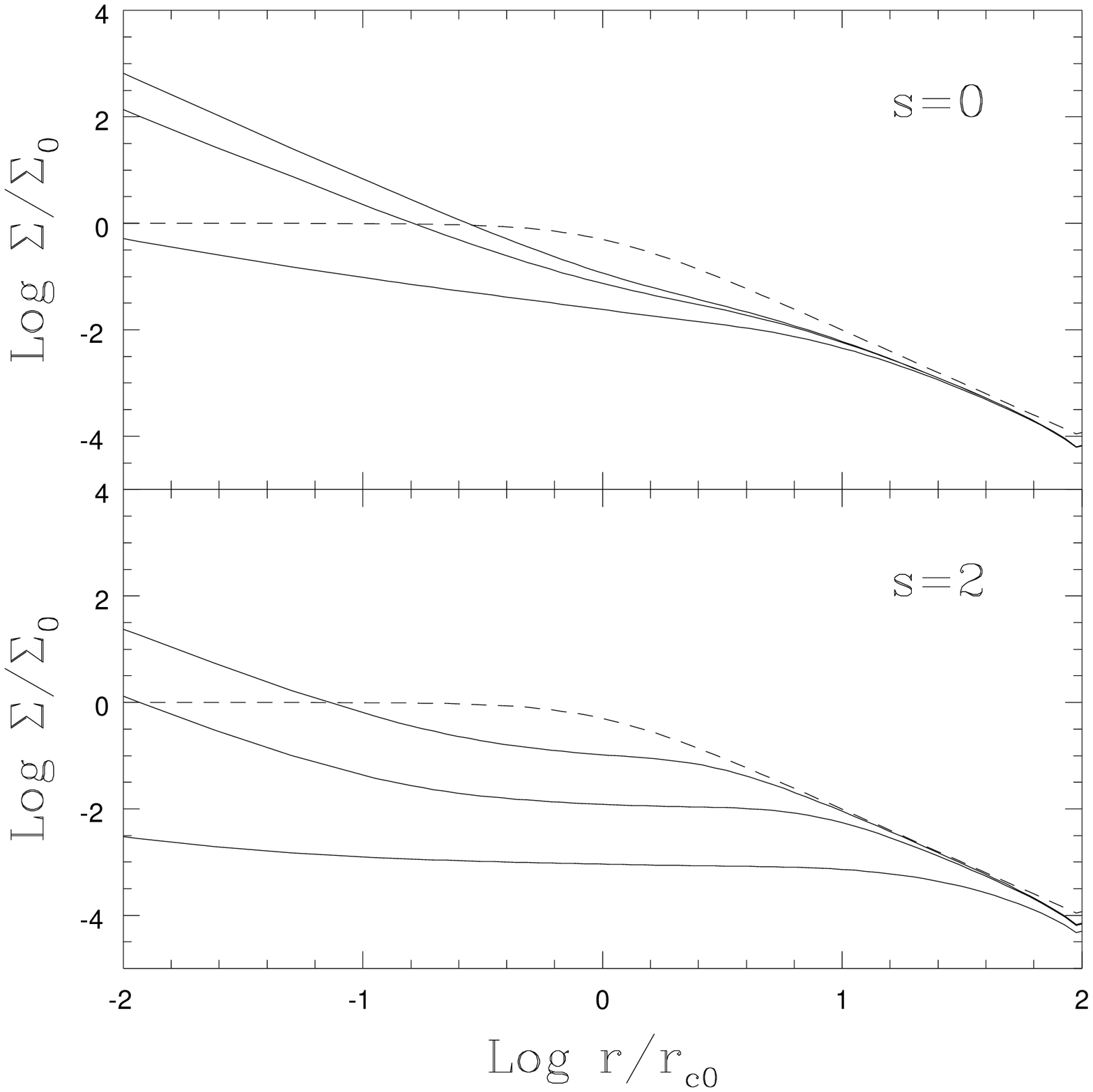}{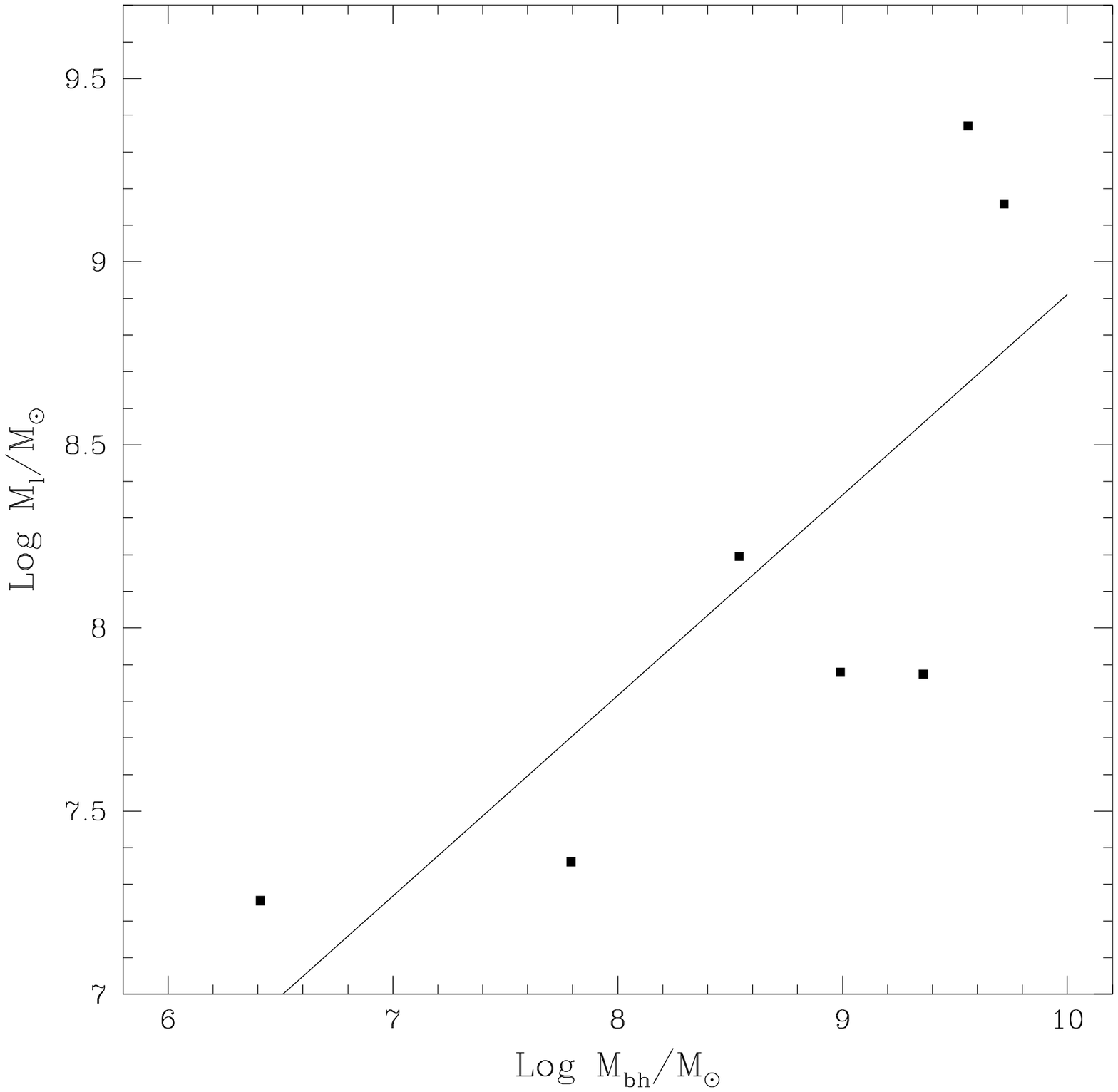}
\caption{ {\bf left panel}:
projected radial density profiles of the GCS at $t=0$ 
(dashed curve) and after a 15 Gyr evolution in presence of an 
initial central 
galactic black hole of mass $10^7$,$10^8$, $10^9$ 
M$_\odot$ (solid curves, from top to bottom, respectively); 
$\Sigma_0$ and $r_{c0}$ are the initial central density and core 
radius of the GCS; $s=0$ and $s=2$ are the slopes of the GCS 
mass function 
$\propto m^{-s}$ (from Capuzzo--Dolcetta \& Tesseri 1997); 
{\bf right panel}: the 
mass lost from the GCS vs. the central galactic black hole mass 
(in M$_\odot$); the straight line is the least-square fit to the data (from 
Capuzzo--Dolcetta \& Donnarumma 2001).
}
\end{figure}

\begin{figure}
\plotone{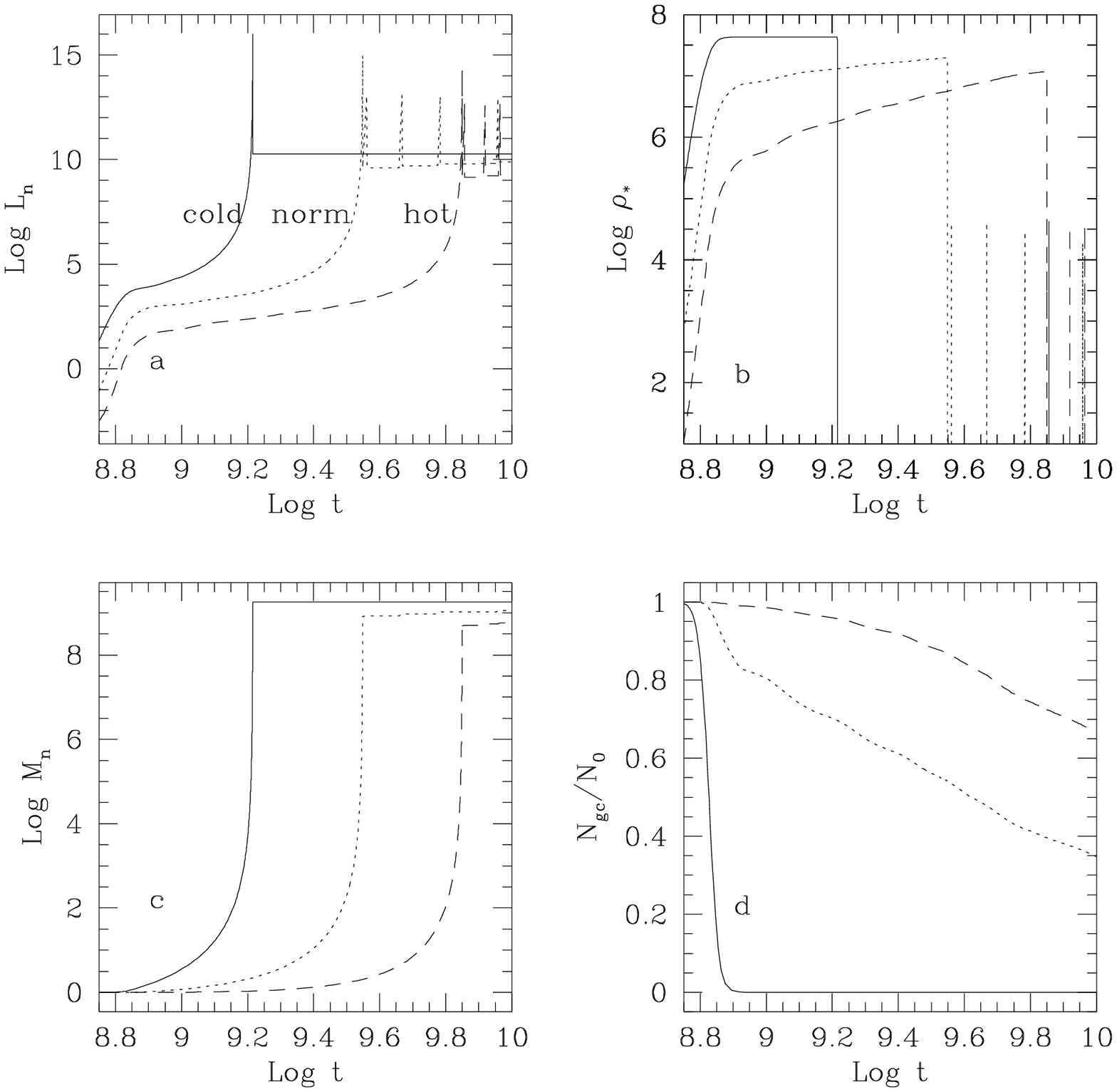}
\caption{
{\bf panels a}: 
time evolution of the nuclear luminosity induced by 
globular cluster merging (time is in $yr$, while 
luminosity is in L$_\odot$); 
{\bf  panel b}: time evolution of the central supercluster mass density 
(in M$_\odot$/pc$^{3})$;
{\bf  panel c}: time evolution of the nucleus mass (in M$_\odot$).
{\bf  panel d}: time evolution of the number of surviving globular clusters 
scaled to its initial value. In all the panels, solid, dotted and dashed curves 
refer to {\it cold}, {\it normal} and {\it hot} models, respectively (see 
text).}
\end{figure}

\par\noindent It results that (in the sample of 17 galaxies examined so far) the 
amount of mass lost from the GCS to the central galactic zones ranges from 25\% 
to 75\% of the initial total mass of the GCS. This corresponds to quantity of 
mass always greater than $10^7$ M$_\odot$, up to the case of the giant 
elliptical M 87 where it is $2.3 \times 10^9$ M$_\odot$. Another point 
corroborating the evolutionary view is the positive correlation of the GCS mass 
lost and the mass of the galactic central black hole estimated for 7 out of the 
17 galaxies studied (Fig. 1). 
\par\noindent These considerations suggest as worth a better investigation of 
the role of the GCS mass loss toward the galactic centre and its possible effect 
as fueling the inner black hole. This has been partly done in Capuzzo--Dolcetta 
(1993),  and deepened and generalized in the forthcoming paper by Capuzzo--
Dolcetta (2001). Here I preliminarly present some of the results that will be 
extensively given and discussed in the latter paper.
\par The details of the model are described elsewhere (for instance
Capuzzo--Dolcetta 2000, Capuzzo--Dolcetta 2001) and they are not worth repeated here. 
\par\noindent The qualitative sketch of what happens is:
\par\noindent
$\bullet $ massive globular clusters moving on box orbits (in triaxial 
potentials) or on low angular momentum orbits (in symmetric potentials) lose 
rather quickly their orbital energy (the time needed depending mostly on their 
initial energy, i.e. on the velocity dispersion of the GCS);
\par\noindent
$\bullet $ after a time of the order of $500$ Myr  many 'decayed' clusters are 
limited to the inner galactic region and merge: the precise modes of this 
merging is still not known (complicated N-body simulations are required) but a 
a {\it supercluster} is likely to be formed;
\par\noindent
$\bullet $ stars of the supercluster buzz around the nucleus, where a pre-
existing black hole captures them with a rate of the order of
\bge 
\dot m = <\rho_*> \sigma <v_*^2>^{1/2},
\ede
where $\rho_*$ and $v_*$ are the density (by mass) and velocity of stars around
the black hole and $\sigma$ is the proper swallowing cross--section;
\par\noindent 
$\bullet $ part of the energy extracted from the gravitational field goes into 
electromagnetic radiation (corresponding to a luminosity $L_n$) and part 
increases the black hole mass $M_n$.

\par\noindent
The mutual feedback between the black hole and the GCS is such that the 
accretion rate onto the black hole increases until its mass is large enough to 
act as an efficient tidal destroyer of lighter clusters that were less 
efficiently braked by dynamical friction; this avoid the residual supercluster 
to grow and, so, limits the accretion process. 
\par Of course, in order the evolutionary scenario described to be relevant all 
the steps described above should occur in sufficiently short times. This is what 
actually happens on the basis of the simplified but realistic model described in
Capuzzo-Dolcetta (2000). A model that it is possible to take as reference is the 
evolution of a GCS composed by $1000$ clusters of the same mass ($2 \times 10^6$ 
M$_\odot$) moving in the parent elliptical galaxy where a black hole of 1 
M$_\odot$ is sited {\it ab initio} at its centre. The initial orbital 
distribution of the GCS is assumed as a box-biased distribution function in the 
form of an isothermal DF multiplied by a function of the orbital angular 
momentum (see Capuzzo-Dolcetta 1993 for details) and is characterized by three 
different choices of the velocity dispersion $\sigma_{GC} =165, 330, 660$ 
km/sec, which correspond to what will be called 'cold', 'normal' and 'hot' 
systems, respectively. The name 'normal' is due to that the corresponding 
velocity dispersion ($330$ km/sec) is the same of the stars of the galaxy. 
\par\noindent As an example of the results, in Fig. 2 the time evolution of some 
relevant quantities is shown. The nuclear luminosity grows up 
to super-Eddington luminosities in all the three cases investigated; later, 
it decreases rapidly, to stabilize around $10^{10}$ L$_\odot$, that is the value 
due to the steady capture of bulge stars by mean of the 'grown up' black hole 
(see Fig. 2a). The final black hole mass is $1.8 \times 10^9$, $1.2 \times 
10^9$ and $7.2 \times 10^8$ M$_\odot$ in the cold, normal and hot case, 
respectively (Fig. 2c).
Panel d of Fig. 2 shows the evolution of the survived component of the GCS: it 
is an essential parameter to rule out too 'cold' GCS, which are more efficient 
in feeding the galactic nucleus but are of course depopulated soon. At $t=15$ 
Gyr the 'cold' GCS is depleted while $\sim 300$ and $\sim 600$ clusters of the 
'normal' and the 'hot' GCSs are still surviving, respectively.

\section{Results}
\par\noindent 
Time evolution of globular cluster systems in galaxies has a relevant role both 
for their own structure and characteristics and for those of the parent galaxy.
\par\noindent
Among the numerous types of feedbacks among GCSs and host galaxies, this short report
briefly discussed the role clusters have played in the initial violent activity of 
galaxies. A self-consistent model whose details are described 
elsewhere was employed; its most interesting  output is the time evolution of the nucleus 
mass and luminosity, which depends, in a non-linear way, on the (local)
 evolution 
of the star mass density around the galactic centre, which, in its turn, depends 
on the evolution of the GCS. 
\par\noindent
Among the various free  parameters in the model, 
the most relevant on the  evolution of the GCS are found to be  
\par\noindent i) the initial number of globular clusters;
\par\noindent ii) the IMF of the GCS;
\par\noindent iii) the orbital 'temperature' of the GCS (the higher the 
temperature the less efficient the evolutionary effects induced by both the 
large- and small-scale mass distribution of the host galaxy);
\par\noindent iv) the initial nucleus mass;
\par\noindent v) the efficiency of conversion of gravitational energy into 
radiation.
\par
The main result presented here is that the combined effects on globular clusters 
of both dynamical friction and tidal interaction with the central compact region 
allow the formation of a central, dense supercluster; if the central stellar 
density reaches sufficiently high values, the supercluster releases mass (and 
gravitational energy) to the nucleus. The time evolution of the nucleus mass and 
luminosity is found to be in the range of many AGNs.  Usually, to a slow 
brightening phase follows a faster one and, later, a slow dimming phase.

\end{document}